# Optical characterization of GaN by N$^+$ implantation into GaAs at elevated temperature

S. Dhara,[a)] P. Magudapathy, R. Kesavamoorthy, S. Kalavathi, and K. G. M. Nair,

Materials Science Division, Indira Gandhi Centre for Atomic Research, Kalpakkam-603 102

G. M. Hsu, L. C. Chen and K. H. Chen[b)]

Centre for Condensed Matter Sciences, National Taiwan University, Taipei-106, Taiwan

K. Santhakumar and T. Soga

Department of Environmental Technology and Urban Planning, Nagoya Institute of Technology, Nagoya 466-8555, Japan

**Abstract :**

Both hexagonal wurtzite and cubic zinc blend GaN phases were synthesized in GaAs by 50 keV N$^+$ implantation at 400 °C and subsequent annealing at 900 °C for 15 min in N$_2$ ambient. Crystallographic structural and Raman scattering studies revealed that GaN phases were grown for fluence above $2 \times 10^{17}$ cm$^{-2}$. Temperature-dependent photoluminescence study showed sharp direct band-to-band transition peak ~3.32 eV at temperature ≤200K. The intermediate bandgap value, with respect to ~3.4 eV for hexagonal and ~3.27 eV for cubic phases of GaN, is an indicative for the formation of mixed hexagonal and cubic phases.

[a)] Corresponding author Email : dhara@igcar.gov.in

[b)] Also affiliated to Institute of Atomic and Molecular Sciences, Academia Sinica, Taipei 106, Taiwan

Recent advances in growth of epitaxial GaN layer and the realization of GaN-based optical devices have inspired a flurry of research activity aimed at exploiting the potential of III-V nitrides as light emitting devices active in the full range of the visible spectrum (blue and UV region in particular) and as electronic devices suitable for high temperature, high power, and high frequency applications.[1] To date, many techniques, including reactive sputtering, chemical vapor deposition, reactive molecular beam epitaxy, and their variations, have been used to grow GaN thin films on a variety of substrates.[2] For the high ionicity in III-V compounds, GaN normally grows in its stable hexagonal phase (h-GaN) with the wurtzite structure, but a metastable cubic phase (c-GaN) of zinc blend structure can also be produced under certain growth conditions.[1-3] Detailed structural characterization is reported for the growth of nanocrystalline GaN phases (cubic and/or hexagonal) by $N^+$ implantation in GaAs [Ref. 4] and GaP [Ref. 5] at elevated temperatures ~400$^0$C. These GaN phases grown by ion implantation technique are useful for the formation of depth selective embedded structures, which will be useful for futuristic opto-electronic applications. However, optical study is generally missing in both the studies.[4,5]

We report here a similar technique of producing GaN, namely, ion-beam synthesis via $N^+$ implantation in a GaAs substrate at an elevated temperature and subsequent annealing treatment. This approach is based on the following considerations i) since GaN is thermodynamically more stable than GaAs (heats of formation $\approx$ -109.5 kJ mol$^{-1}$ and -81.5 kJ mol$^{-1}$, respectively),[6] introduction of N atoms in GaAs is expected to cause N–As exchange and result in GaN formation, ii) the replacement of As by N atoms is facilitated by the fact that As is more volatile than Ga and tends to escape from GaAs upon thermal annealing, and iii) ion implantation is a technique that allows for substantial introduction of N atoms in GaAs, far beyond the solid solubility limit. Glancing-angle x-ray diffraction (GIXRD) and Raman scattering techniques were used for the identification of the GaN phases. Optical properties of the embedded GaN structure were studied in details, for the first time.

50 keV $N^+$ implantation in crystalline GaAs(100) substrates at 400 $^0$C was performed using gaseous ion source 150 kV ion implanter. The reason for choosing implantation at elevated temperature is to minimize defects introduced during energetic implantation process, which in a turn may help in achieving reasonably good crystallinity in the grown sample. A low beam current of ~ 4 μA cm$^{-2}$ was used to achieve fluences in the range of $1\times10^{17} - 4\times10^{17}$ cm$^{-2}$. As calculated from the SRIM code,[7] the range of 50-keV $N^+$ in GaAs is ~119 nm. Implanted samples were annealed at a temperature of 900 $^0$C for 15 minutes in ultra high pure nitrogen ambient. The evolution of phases created by ion bombardment was studied using grazing incidence X-ray diffraction (GIXRD; STOE Diffractometer). All the measurements were carried out with an angle of incidence, ω= 0.3$^o$ for X-rays. Raman spectra (400 - 790 cm$^{-1}$) of the as-grown and high temperature $N^+$ implanted samples were recorded in back scattering geometry using 514.5 nm Ar$^+$ laser line. A temperature-dependent photoluminescence (PL) study using He-Cd laser (excitation wavelength of 325 nm) with power fixed at 40 mW was performed to measure optical properties. A power dependence PL study at lowest temperature (20K) was also performed in order to confirm the direct bang-gap peak.

The GIXRD patterns of the unimplanted and annealed samples implanted at $4\times10^{17}$ cm$^{-2}$ are shown in Fig. 1. The peaks at 2θ values of ~ 45.54$^0$, and 53.78$^0$ correspond to reflections from (220) and (311) planes of cubic-GaAs substrates.[8] Intensities from planes other than (100) plane of crystalline GaAs are possible in the ω-2θ scan. The GIXRD pattern for the annealed sample implanted with a fluence of $4\times10^{17}$ cm$^{-2}$ shows (Fig. 1) peaks at 2θ value of ~ 44.08$^0$ corresponding to (200) plane of c-GaN phase.[8] We performed a slow scan for the implanted post-annealed sample in the range of 2θ values of ~57$^0$-75$^0$ for detailed structural analysis. Peaks at 2θ values of ~59.6$^0$, 62.5$^0$ corresponding c-GaN and ~64.4$^0$, 67.8$^0$, 73.5$^0$ corresponding to h-GaN phases.[8] The corresponding planes are indicated in the figure. The observed 2θ values for the peaks corresponding to cubic and hexagonal phases are in the higher side than that reported for the bulk sample,[8] indicating shrinkage in lattice constants and hence presences of compressive strain in the embedded precipitates of GaN in GaAs. Generally low intensities of the diffraction peaks indicate formation of low amount of crystalline phases in the embedded structure. We failed to identify any appreciable phase formation for the annealed samples implanted at fluences below $4 \times 10^{17}$ cm$^{-2}$ (not shown in figure).

Raman scattering study of the annealed sample implanted at fluences below $4 \times 10^{17}$ cm$^{-2}$ did not show prominent features of GaN phase formation. Figure 2 shows Raman peaks around 450 cm$^{-1}$, 539 cm$^{-1}$, 729 cm$^{-1}$ and 755 cm$^{-1}$ which correspond to Zone Boundary (ZB) mode, transverse optical mode of $A_1$(TO), longitudinal optical modes of $A_1$(LO) and $E_1$(LO), respectively of h-GaN phase reported in nanowire.[9] Presence of ZB peak around 450 cm$^{-1}$ indicates finite size effect in the embedded structure of GaN in GaAs. The Raman peaks are shifted on higher side than those observed for the bulk value,[10] indicating compressive strain in the embedded system.[11] Peak around 775 cm$^{-1}$ corresponds to LO mode of c-GaN phase, which is also shifted on higher side than that reported for thin film sample.[12] Presence of both cubic and hexagonal phases with compressive strain in the embedded structure support the observation of crystallographic structural data (Fig. 1). The low intensities of the Raman peaks indicate formation of low amount of crystalline phases in the embedded structure.

Optical properties in our temperature-dependent PL study showed (Fig. 3) a sharp peak at ~3.32 eV at temperatures ≤ 200K and designated as direct band-to-band transition peak. The expected blue shift of the band-to-band transition peak with decreasing temperature is not observed. This may be due to the fact that the blue shift of PL peak with decreasing temperature as well as due to the compressive strain,[11] as observed from crystallographic and Raman studies in the embedded system, is compensated with difference in thermal expansion coefficient of GaN ($7.75 \times 10^{-6}$ K$^{-1}$) and GaAs ($5.7 \times 10^{-6}$ K$^{-1}$) producing tensile strain in the system. Small intensity of the band-to-band transition peak may be due to the formation of small amount crystalline GaN phases in embedded structure, as also reflected in our crystallographic structural and Raman scattering studies. The confirmation for direct band-to-band transition peak at ~3.32 eV comes from our power-dependent PL study (Fig. 4). The variation of PL intensities at ~3.32 eV with power shows a linear fitting (inset Fig. 4), proving that it is indeed a direct band-to-band transition peak for grown GaN phase. The intermediated band-gap value of ~3.32 eV to with respect to ~3.4 eV for h-GaN and ~3.27 eV for c-GaN, is an indicative for the formation of mixed hexagonal and cubic phases,[13] as also indicated in our structural studies using GIXRD and Raman scattering. PL spectrum at room temperature shows a broad band around ~2.6 eV, which may be due to nitrogen vacancy ($V_N$) related defects,[14] as short annealing time in the embedded structure bound to have nitrogen deficiency. Short annealing time was preferred to avoid major depletion of As in GaAs lattice and to get rid of complexities in structures of grown GaN phase. The broad peak with possible origin of $V_N$ related defect is narrowed and shifted to ~3.0 eV at lower temperatures. This may be due to stabilization of defects at lower temperature.

In conclusion, both the wurtzite and zinc blend phases of GaN was grown by a two-step technique comprising of $N^+$ beam implantation in GaAs at an elevated temperature and subsequent high temperature annealing treatment in $N_2$ ambient. GaN phase is grown at a high fluence of $4 \times 10^{17}$ cm$^{-2}$. Presence of both hexagonal and cubic phases of GaN was confirmed by crystallographic structural and Raman scattering studies. Optical studies showed a bandgap at ~3.32 eV corresponding to mixed hexagonal and cubic GaN when measured ≤ 200K.

We acknowledge J. C. George, MSD for the heater arrangement. We also thank C. S. Sundar, MSD for his encouragement in pursuing this work.

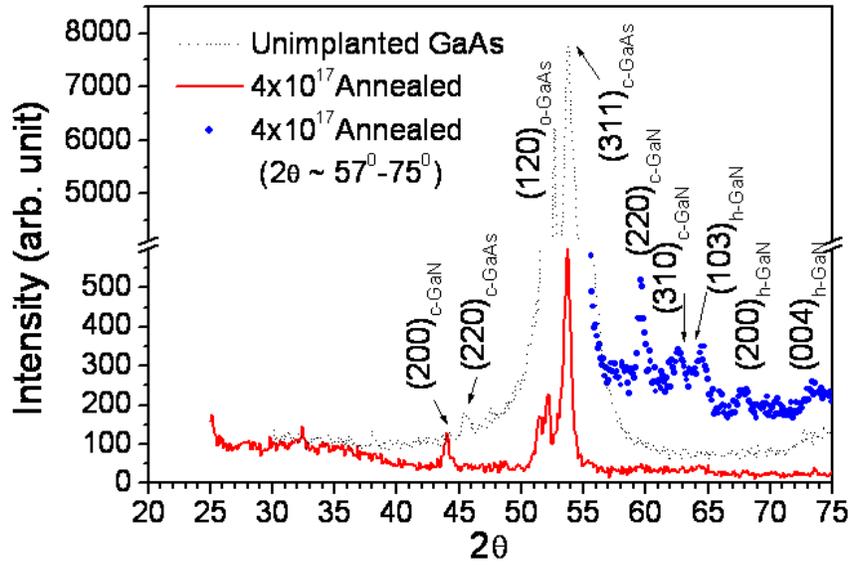

Fig. 1. (Color online) GIXRD study of the unimplanted, and post-implanted annealed samples implanted at fluence of $4\times10^{17}$cm$^{-2}$. Slow scan data ($2\theta \sim 57$-$77^0$) for the implanted post-annealed sample is shown as scattered points to identify number of GaN phases.

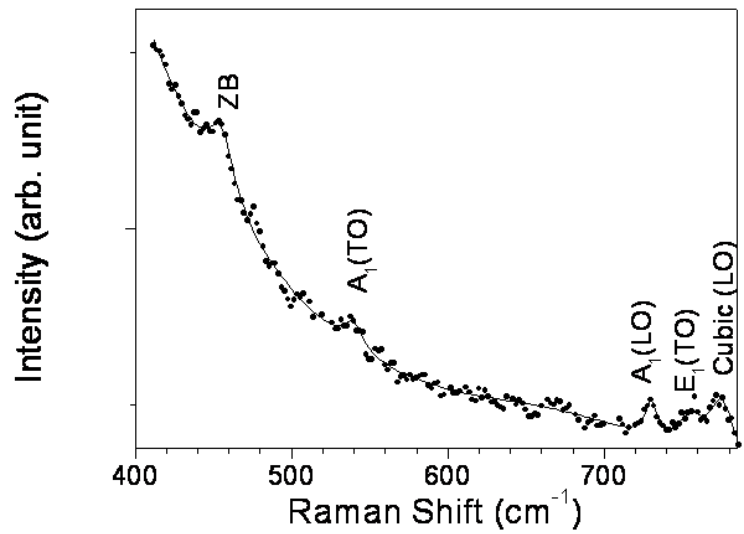

Fig. 2. Raman scattering study of annealed sample implanted at fluences of $4\times10^{17}$ cm$^{-2}$.

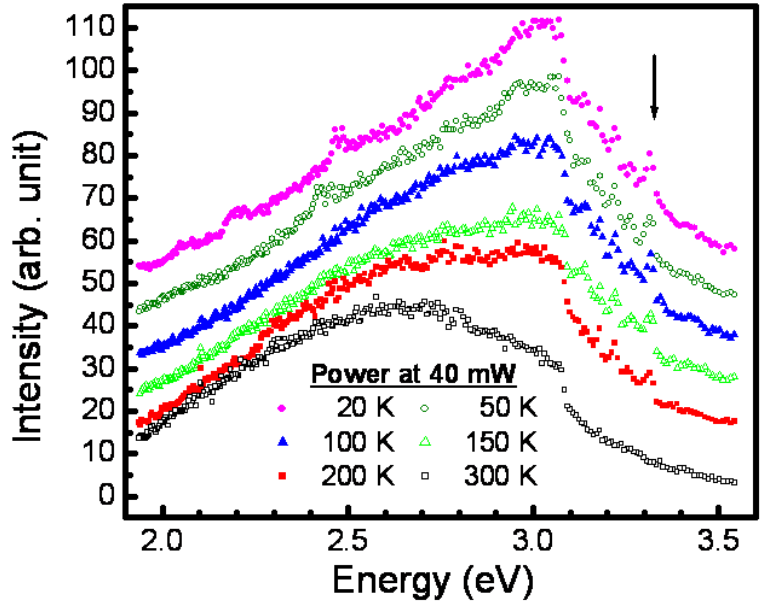

Fig. 3. (Color online) Temperature-dependent PL spectra for the post-annealed GaN implanted with a fluence of $4\times10^{17}$ cm$^{-2}$ showing presence of band-to band transition peak ~3.32 eV (indicated with an arrow). Spectra are shifted vertically for clarity.

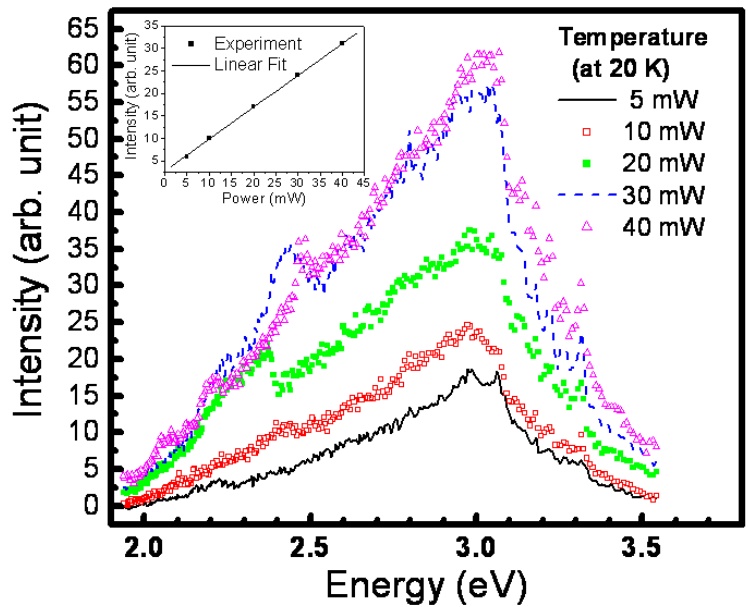

Fig. 4. (Color online) Excitation power-dependent PL spectra for the post-annealed GaN implanted with a fluence of $4\times10^{17}$ cm$^{-2}$. Inset shows variation of PL peak (at ~3.32 eV) intensities with the excitation power and the linear fitting.